\documentclass[10pt,a4paper,twoside]{article}

\newenvironment{resum}{\begin{quote}\small}{\end{quote}}

\newcommand{\bfsf}[1]{\textsf{\textbf{#1}}}

\def\ii{\'{\char'20}}
\def\beq{\begin{equation}}
\def\eeq{\end{equation}}
\def\bea{\begin{eqnarray}}
\def\eea{\end{eqnarray}}
\def\dGG{\dot{G}/G}
\def\fdGG{\frac{\dot{G}}{G}}
\def\daa{\dot{\alpha}/\alpha}
\def\fdaa{\frac{\dot{\alpha}}{\alpha}}

\def\fdRR{\frac{\dot{R}}{R}}
\def\lapprox{\hbox{\lower .8ex\hbox{$\,\buildrel < \over\sim\,$}}}
\def\gapprox{\hbox{\lower .8ex\hbox{$\,\buildrel > \over\sim\,$}}}

\begin{document}

\thispagestyle{plain}		

\begin{center}


{\LARGE\bfsf{Time  variation of the  gravitational and  fine structure
constants in models with extra dimensions}}\footnote{Talk presented by
Yu.A.  Kubyshin at  the  Spanish Relativity  Meeting ERE-2002  (Ma\`o,
Spain, September 22-24, 2002)}


\textbf{E. Garc{\ii}a--Berro}\dag \ddag,
\textbf{Yu.A. Kubyshin}\dag \P and
\textbf{P. Lor\'en--Aguilar}\dag


\dag\textsl{Universitat Polit\`ecnica de Catalunya, Spain.} \\
\ddag\textsl{Institut d'Estudis Espacials de Catalunya, Spain.} \\
\P\textsl{Institute  of  Nuclear  Physics,  Moscow  State  University,
Russia.}

\end{center}


\begin{resum}
We  derive  formulae  for  the  time variation  of  the  gravitational
``constant'' and of the  fine structure ``constant'' in various models
with  extra   dimensions  and  analyze  their   consistency  with  the
observational data.
\end{resum}


\section{Introduction}

Physical  theories  contain   certain  parameters  characterizing  the
strength  of the  interaction which  are  assumed to  be constant  and
fundamental.   Examples  of  such  parameters  are  the  gravitational
constant  $G$, which characterizes  the gravitational  interaction, or
the fine  structure constant $\alpha$ determining the  strength of the
electromagnetic   interaction.    However,   there  are   experimental
evidences of  the temporal variation  of some of these  constants.  In
particular,  it has  been  recently reported  \cite{Mur00} a  positive
detection  of  the variability  of  the  fine  structure constant,  by
comparing  quasar  absorption lines  at  $z\simeq  3$ with  laboratory
spectra.  Results on the  time variation of the gravitational constant
$G$  \cite{Uzan02} usually  yield experimental  bounds $|  \dot  G/G |
\lapprox  10^{-11}  \;   \mbox{yr}^{-1}$.   Finally,  more  conclusive
results,  based on  the analysis  of the  Hubble diagram  for  type Ia
supernovae were  reported in Ref.  \cite{Gazt01}.  On  the other hand,
there are a  few theoretical schemes which predict  such variations of
the  fundamental   constants  ---   see  Ref.   \cite{Uzan02}   for  a
comprehensive  review on  this subject.   Here we  study  models which
appear  as  a  low-energy  limit  of some  ``fundamental''  theory  in
$(4+d)$-dimensions.  Our interest in such theories is motivated by the
fact  that they provide  a natural  and self-consistent  framework for
such variations.   Previous studies  on this subject  can be  found in
\cite{Mar84,Xdim-var}.   In what  follows  we study  three classes  of
multidimensional  models   and  derive  relationships   for  the  time
variation of $\alpha$  and $G$.  These relationships will  be used for
establishing predictions  for the time variation  of the gravitational
constant.

\section{Theories with extra dimensions}

T.  Kaluza  and  O.   Klein  in  their  pioneerings  papers  \cite{KK}
discovered that  the zero-mode sector  of the Einstein gravity  in the
five-dimensional space-time $M^{4} \times  S^{1}$ --- where $M^{4}$ is
the Minkowski space-time  and $S^{1}$ is the circle  --- is equivalent
to the  Einstein gravity and Maxwell electrodynamics  in $M^{4}$.  The
relation between  the parameters  of the four-dimensional  theory, $G$
and $\alpha$, and the constant $\hat{G}_{(5)}$ of the five-dimensional
theory is given by the following reduction formula:
\beq
   G = \frac{\hat{G}_{(5)}}{2\pi R}, 
   \; \; \; \alpha =  \frac{\hat{G}_{(5)}}{2 \pi R^{3}}.     
\label{G-G5-red}
\eeq
The  construction was  later  generalized \cite{KK1}  to more  general
spaces $M^{4} \times K_{(d)}$  (see Refs. \cite{Duff} for reviews). In
this  case the  $(4+d)$-dimensional  Einstein gravity  reduces to  the
gravity and Yang-Mills theory in  four dimensions with the gauge group
being  the isotropy  group  of  the space  of  extra dimensions.   The
reduction formulae are
\beq
G = \frac{\hat{G}_{(4+d)}}{V_{(d)}} \propto 
\frac{\hat{G}_{(4+d)}}{R^{d}}, 
\; \; \; 
\alpha = \kappa \frac{\hat{G}_{(4+d)}}{R^{2} V_{(d)}},   
         \label{G-Gd-red} 
\eeq
where $\kappa$ is some numerical  factor which depends on the specific
model and  $V_{(d)}$ is the volume  of the space  of extra dimensions.
From these relations  one easily concludes that $\dot G/G  = - d (\dot
R/R)$, and $\dot\alpha/\alpha = - (d + 2) (\dot R/R)$. Consequently,
\beq
\fdaa = \frac{d+2}{d} \fdGG.  
\label{da-dG-KK}
\eeq
Let  us consider  now an  Einstein-Yang-Mills theory  formulated  in a
$(4+d)$-dimensional  space-time of the  form $M_{(4)}  \times K_{(d)}$
\cite{CSDR,CSDR-rev}.  The theory  includes gravity and the Yang-Mills
field, and its action is given by
\beq
S = \int_{M_{(4)} \times K_{(d)}} d^{4+d} \hat{x} \sqrt{-\hat{g}} 
\left[\frac{1}{16 \pi \hat{G}_{(4+d)}} {\cal R}^{(4+d)} +
  \frac{1}{4 \hat{g}_{(4+d)}^{2}} Tr \hat{F}_{MN} \hat{F}^{MN} \right] ,
   \label{D-EYM}
\eeq
where,   as   above,   $\hat{G}_{(4+d)}$   is   the   multidimensional
gravitational constant  and $\hat{g}_{(4+d)}$ is  the multidimensional
gauge  coupling.  Both  are  supposed  to be  constant  in  time.  The
dimensionally  reduced  theory  includes  the  Einstein  gravity,  the
four-di\-men\-si\-o\-nal gauge fields and scalar fields.  The explicit
form of the  dimensionally reduced theory depends on  the topology and
geometry  of the space  of extra  dimensions and  the multidimensional
gauge group.  The four dimensional couplings are given by
\[
G = \frac{\hat{G}_{(4+d)}}{V_{(d)}} \propto \frac{G_{(4+d)}}{R^{d}},
\; \; \; \alpha = \frac{\hat{\alpha}_{(4+d)}}{V_{(d)}} 
 \propto \frac{\hat{\alpha}_{(4+d)}}{R^{d}}. 
\]

As a consequence  the time variations of $G$  and $\alpha$ are related
as follows:
 
\beq
    \fdaa = \fdGG.      \label{daG-EYM}
\eeq

Finally, let  us consider the  Randall-Sundrum (RS) model  proposed in
Refs.   \cite{RS1}  (see  \cite{RS-review,Rub01}  for reviews  on  the
subject). It describes the  five-dimensional Einstein gravity with the
cosmological    constant    in    the   space-time    $M^{4}    \times
S^{1}/Z_{2}$. Here  $S^{1}/Z_{2}$ is the orbifold,  the space obtained
from the circle $S^{1}=\{y| 0 \leq y < 2\pi R \}$ of radius $R$ by the
identification $y \cong  (-y)$. There are two 3-branes  located at the
fixed points  $y=0$ and $y=\pi R$  of the orbifold,  one with positive
brane tension $(\sigma)$ and the other with the negative brane tension
$(-\sigma)$.   The gravity  propagates in  the  five-dimensional bulk,
while   matter  fields   are   supposed  to   be   localized  on   the
branes. Usually the  brane with negative tension is  located at $y=\pi
R$ and identified with our physical 3-space.  The RS model provides an
elegant  geometrical solution  to the  hierarchy problem  and predicts
some physical effects  which, in principle, can be  observed in future
collider  experiments.    To  find  the  reduction   formula  for  the
gravitational constant we transform first to the coordinates which are
Galilean on the physical  brane; that is, on the brane at  $y = \pi R$
\cite{BKSV02}.   In these  coordinates  the Planck  mass is  expressed
through the fundamental scale $M_{(5)}$ of the five-dimensional theory
(the five-dimensional Planck mass) as follows \cite{Rub01,BKSV02}
\beq  
M_{\rm Pl}^{2}  =  \frac{M_{(5)}^{3}}{k}   
\left[  {\rm e}^{2\pi  k  R}  - 1\right],  
\label{RS-Mpl-M5} 
\eeq
where $k$  is a parameter related  to the brane  tension $\sigma$.  It
turns  out that  the fundamental  scale and  $k$ must  be of  order of
$1\;$TeV. For the hierarchy problem to  be solved the value $k R$ must
be approximately  $k R \approx  11 - 12$.  From  Eq. (\ref{RS-Mpl-M5})
one easily gets that

\beq
G = \frac{k}{16 \pi M_{(5)}^{3}} \frac{1}{{\rm e}^{2\pi k R} - 1}. 
\label{RS-G-M5}
\eeq

In the original formulation \cite{RS1}  all the fields of the Standard
Model were localized on the negative  tension brane. It is easy to see
that in  this case the  model does not  provide any mechanism  for the
time  variation for  the fine  structure constant  $\alpha$.  Extended
versions  of the  RS  model  were also  considered  in the  literature
\cite{DHR99,DHR01}.  One  of the possibilities  is to allow  gauge and
some  matter  fields to  propagate  in the  bulk.   In  this case  the
four-dimensional  gauge  coupling  on  the  brane is  related  to  the
five-dimensional   one    as   follows   \cite{DHR01}:    $g_{(4)}   =
\hat{g}_{(5)}/  {\sqrt{2\pi R}}$.  Now  suppose that  the size  of the
space of  extra dimensions varies with  time.  This leads  to the time
variation of the gravitational and fine structure parameters given by

\beq
 \fdGG = - (2 \pi k R) \frac{1}{1-e^{-2\pi k R}} \fdRR 
  \approx - (2 \pi k R) \fdRR,  \; \; \; 
 \fdaa = - \fdRR. 
\label{RS-dGG}
\eeq  
From these two relations we get 
\beq
\fdaa \approx \frac{1}{2 \pi k R} \fdGG.    
\label{RS-daa-dGG}
\eeq

To  summarize, we  have analyzed  three classes  of models  with extra
dimensions,   namely  the   ``classical''  Kaluza-Klein   models,  the
Einstein-Yang-Mills models  and the extended  version of the  RS model
with gauge fields  propagating in the bulk. We  have obtained that for
these three cases there is a relationship between the respective rates
of variations of $\alpha$ and $G$:
\beq 
\fdaa = \beta (R) \fdGG, \label{daa-dGG} 
\eeq
where $\beta(R)$ depends on the  adopted model.  The factor $\beta$ is
$\sim 1$  for the first and  second class of models.   For the RS-type
model,  since $k  R  \approx  11-12$ the  parameter  $\beta$ is  $\sim
10^{-2}$.  It is important to realize that in all the cases $\beta (R)
> 0$.  Consequently, the time variations  $\daa$ and $\dGG$ are of the
same sign. We emphasize that  this feature appears to be quite generic
and  model  independent  for   theories  with  extra  dimensions,  and
therefore our result  is rather robust.  

Let us  see some  implications of our  result.  The recent  results of
\cite{Mur00}   yield   $\Delta   \alpha/\alpha  \equiv   (\alpha_z   -
\alpha_0)/\alpha_0  \sim  -  10^{-5}$  for  $  0.5  \leq  z  \leq  2$.
Assuming,  as  it  is usual,  a  constant  rate  and using  a  typical
look-back time $\Delta  t \approx 8 \cdot 10^{9}  \;$yr (at $z=1$) one
obtains $\dot{G}/G \sim + 10^{-15} \; \; \mbox{yr}^{-1}$ for the first
two models  and $\dot{G}/G \sim  + 10^{-13} \; \;  \mbox{yr}^{-1}$ for
the RS model.  We see that  in both cases the predicted time variation
for the gravitational constant is positive.

Let us  examine now the experimental  bounds on the  time variation of
$G$. From  the best  fit to  the Hubble diagram  of SNIa  at $1\sigma$
C.L.   \cite{Gazt01}  bounds   on   $\dGG$  for   various  values   of
$(\Omega_{M},\Omega_{\Lambda})$  can be obtained.   Here we  present a
few examples of them for $z = 0.5$:

\smallskip
\begin{tabular}{ll}
$\Omega_{M}=1.0, \; \; \Omega_{\Lambda}=0.0 \; \; \; \;$ & 
$-3.0 \cdot 10^{-11} \; \mbox{yr}^{-1} < \dot G/G < 
-0.8 \cdot 10^{-11} \; \mbox{yr}^{-1}$,  \\
$\Omega_{M}=0.3, \; \; \Omega_{\Lambda}=0.7 \; \; \; \;$ & 
$-0.8 \cdot 10^{-11} \; \mbox{yr}^{-1} < \dot G/G < 
+ 1.4 \cdot 10^{-11} \; \mbox{yr}^{-1}$,  \\  
$\Omega_{M}=0.5, \; \; \Omega_{\Lambda}=0 \; \; \; \;$ & 
$-2.0 \cdot 10^{-11} \; \mbox{yr}^{-1} < \dot G/G < 
-1.0 \cdot 10^{-12} \; \mbox{yr}^{-1}$.  \\
\end{tabular}
\smallskip
 
\noindent The comparison of these estimates with our predictions shows
that for  a wide  range of $(\Omega_{M},\Omega_{\Lambda})$  the models
with extra  dimensions considered here  are at odds with  the existing
experimental bounds on the time variation of $G$ and $\alpha$.

Note,   however,  that  our   analysis  relies   on  the   results  of
\cite{Mur00},  which have  been challenged  \cite{Uzan02} and  need an
independent confirmation.  Also, we expect that future improvements in
the experimental  data and/or new experiments will  give better bounds
on $\dGG$ and at least will  determine its sign.  There exists as well
the   possibility  that  the   discrepancy  between   our  theoretical
predictions and  the experimental bounds  quoted above is of  a deeper
nature  and  points towards  some  drawbacks  of the  multidimensional
models  considered here or  even of  the multidimensional  approach as
such. Of course, questioning the applicability of the multidimensional
approach  to the  description  of the  fundamental interactions  needs
further studies.

\vspace{0.2cm}

\noindent  {\bf  Acknowledgements.}    The  authors  are  grateful  to
J. Garriga,  E. Gazta\~naga, J.~-Ph.~Uzan and E. Verdaguer  for useful
discussions and suggestions.  The work of Y.~Kubyshin was supported by
the grant  CERN/FIS/43666/2001 and  RFBR grant 02-02-16444.  This work
has  been partially  supported  by the  MCYT  grants AYA2000-1785  and
AYA2002-04094-C03-02/03 and by the CIRIT.


\end{document}